\documentclass[acmsmall]{acmart}
\usepackage[utf8]{inputenc} 
\usepackage{wrapfig}
\usepackage{comment}
\usepackage{enumitem}
\usepackage{url}
\usepackage{blindtext}
\usepackage{wrapfig,lipsum,booktabs}
\usepackage{graphics}
\usepackage{subfig}
\AtBeginDocument{%
  \providecommand\BibTeX{{%
    \normalfont B\kern-0.5em{\scshape i\kern-0.25em b}\kern-0.8em\TeX}}}

\copyrightyear{2022}
\acmYear{2022}
\setcopyright{rightsretained}
\acmJournal{PACMHCI}
\acmYear{2022} \acmVolume{6} \acmNumber{CSCW2} \acmArticle{527} \acmMonth{11} \acmPrice{}\acmDOI{10.1145/3555640}
\received{January 2022}
\received[revised]{April 2022}
\received[accepted]{August 2022}

\usepackage{etoolbox}
\makeatletter
\patchcmd{\maketitle}{\@copyrightpermission}{
   \begin{minipage}{0.2\columnwidth}
     \href{https://creativecommons.org/licenses/by/4.0/}{\includegraphics[width=0.90\textwidth]{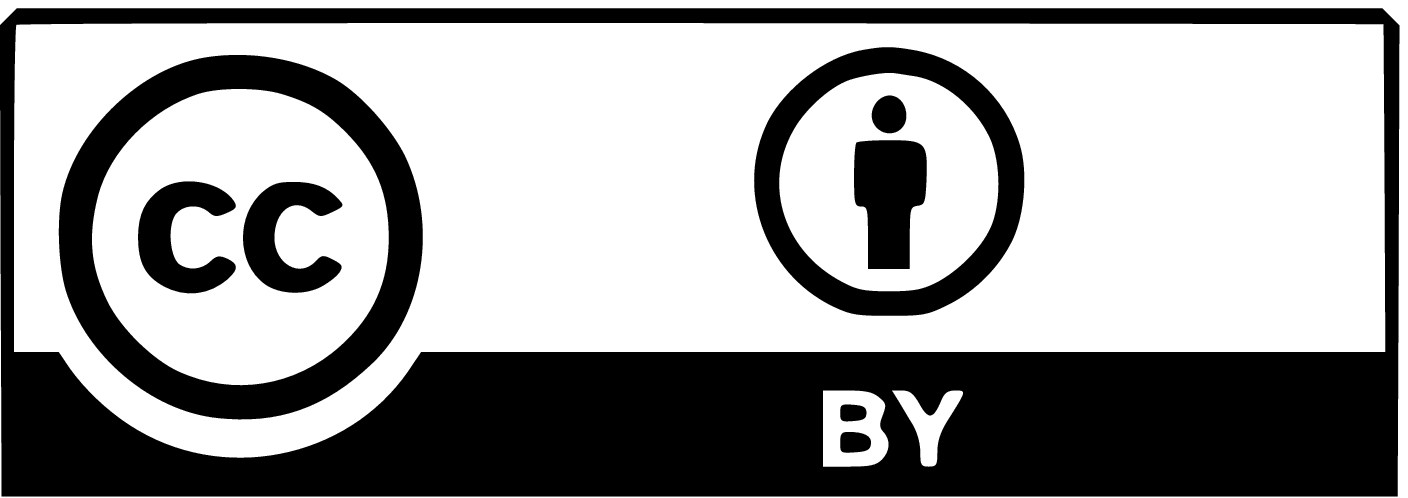}}
   \end{minipage}\hfill
   \begin{minipage}{0.8\columnwidth}
     \href{https://creativecommons.org/licenses/by/4.0/}{This work is licensed under a Creative Commons Attribution International 4.0 License.}
   \end{minipage}

   \vspace{5pt}
}{}{}

\makeatother

\begin{document}

    \title[Modeling Motivational Interviewing Strategies]{Modeling Motivational Interviewing Strategies On An Online Peer-to-Peer Counseling Platform}


 \author{Raj Sanjay Shah}
 \authornote{Both authors contributed equally to this research.}
 \affiliation{%
  \institution{Georgia Institute of Technology}
  \country{USA}}
 \email{rajsanjayshah@gatech.edu}
 

\author{Faye Holt}
 \affiliation{%
  \institution{Georgia Institute of Technology}
  \country{USA}}
 \email{mholt9@gatech.edu}
\authornotemark[1]

 \author{Shirley Anugrah Hayati}
 \affiliation{%
   \institution{Georgia Institute of Technology}
   \country{USA}}
 \email{shirley@gatech.edu}

 \author{Aastha Agarwal}
 \affiliation{%
   \institution{Georgia Institute of Technology}
   \country{USA}}
  \email{aagrawal319@gatech.edu}
 
 \author{Yi-Chia Wang}
 \affiliation{%
  \institution{Independent Researcher}
   \country{USA}}
  \email{yichia.wang@gmail.com}

 \author{Robert E. Kraut}
 \affiliation{%
   \institution{Carnegie Mellon University}
   \country{USA}}
 \email{robert.kraut@cmu.edu}

\author{Diyi Yang}
\affiliation{%
   \institution{Stanford University}
   \country{USA}}
  \email{diyiy@stanford.edu}

\renewcommand{\shortauthors}{Raj Sanjay Shah et al.} 


\begin{abstract}
Millions of people participate in online peer-to-peer support sessions, yet there has been little prior research on systematic psychology-based evaluations of fine-grained peer-counselor behavior in relation to client satisfaction. This paper seeks to bridge this gap by mapping peer-counselor chat-messages to motivational interviewing (MI) techniques. We annotate 14,797 utterances from 734 chat conversations using 17 MI techniques and introduce four new interviewing codes such as ``\emph{chit-chat}'' and ``\emph{inappropriate}'' to account for the unique conversational patterns observed on online platforms. We automate the process of labeling peer-counselor responses to MI techniques by fine-tuning large domain-specific language models and then use these automated measures to investigate the behavior of the peer counselors via correlational studies. Specifically, we study the impact of MI techniques on the conversation ratings to investigate the techniques that predict clients' satisfaction with their counseling sessions. When counselors use techniques such as reflection and affirmation, clients are more satisfied. Examining volunteer counselors' change in usage of techniques suggest that counselors learn to use more introduction and open questions as they gain experience.  
This work provides a deeper understanding of the use of motivational interviewing techniques on peer-to-peer counselor platforms and sheds light on how to build better training programs for volunteer counselors on online platforms.
\end{abstract}



\maketitle

\section{Introduction}
\label{sec:introduction}
Peer support occurs when non-professionals give and receive emotional support, shared knowledge, and shared social or practical help to others suffering from some problem  \cite{peer_support_definition}. Prior research indicates that peer support services can be an integral part of mental health recovery \cite{samhsa_substance_abuse_and_mental_health_services_administration_tip_2013} and benefits support seekers \cite{intro_online_support_works}, even though  most peer counselors have not received professional training \cite{peer_sup_intro}. Online peer-to-peer support communities for mental health, such as Talklife \cite{talklife}, 7 Cups of Tea \cite{7cups}, and MellowTalk \cite{mellowtalk}, have become more popular because they are easy to access and less costly than support from professional therapists \cite{intro_prevalence_of_online_mh}. The number of support seekers using online peer counseling platforms and online mental health communities (OMHCs) has grown significantly in the last few years \cite{online_person_centered} and has seen a surge in demand due to the COVID-19 pandemic \cite{ijerph18052361}. To address the shift of many counseling service to peer-to-peer online formats, there is an increased need to explore how counseling can best adapt to online media. To fill this gap, this paper focuses on one such peer-to-peer counseling platform, using 7 Cups of Tea (7Cups) as a case study, to understand how peer-counselors behave, with a future goal of creating successful training programs for peer-counselors.

On 7Cups, volunteer counselors receive a short training session based on the client-centered guidelines of active listening and motivational interviewing \cite{YAO_CMU}. Motivational interviewing (MI) is a set of client-centric counseling techniques used by therapists to implicitly direct patients to make lifestyle changes or perform actions towards their recovery
\cite{MI_original}. MI focuses on collaboration between the volunteer counselors (known as listeners in 7Cups) and support seekers (known as members), in which the counselors interact with clients using specific skills (or techniques) \cite{MITI, MI_original, MISC, MITI_4}. Being able to identify and measure these skills at scale allows for a detailed analysis of strategies used in counseling sessions. 
Lundal et al. \cite{effective_MI_1} have shown MI to be effective in bringing positive behavioral change for support seekers through a large-scale meta-analysis of 119 studies. 
Magill et al. \cite{magill_MI_causal_model} have shown the efficacy of MI-consistent techniques in promoting client language in support of behavior change. Conversely, MI-inconsistent techniques were shown to promote client language \emph{against} behavior change, which is correlated with negative outcomes for clients \cite{magill_MI_causal_model}. 


We utilize client satisfaction as an outcome metric to explore the value of individual MI techniques, following prior work \cite{intro_prev_online_mh_papers, client_satis_1, intro_online_support_works}. Although therapeutic alliance is a traditional metric used in many studies of the success of OMHCs \cite{ther_alliance, sharf_ther_alliance, martin_ther_alliance}, due to the complexity of measuring it \cite{ardito}, client satisfaction provides a useful proxy readily available as an outcome measure in many OMHCs. Additionally, client satisfaction is linked to therapeutic alliance, which is generally associated with retention and behavioral change \cite{ther_alliance, martin_ther_alliance, sharf_ther_alliance}. On 7Cups, clients can choose a rating based on a 5-point Likert scale from 1 (worst) to 5 (excellent) to express their satisfaction of their conversations with listeners.
In this work, we explore MI usage at a large-scale within an online medium. We are especially interested in MI techniques that predict client satisfaction and are correlated with client change talk from prior literature. 
The focus of our paper, therefore, is to understand which MI techniques that counselors use in  OMHCs that lead to client satisfaction. 
Thus, we address the following two research questions:
\begin{enumerate}
    \item[\textbf{RQ1:}] Which motivational interviewing techniques predict more satisfying conversations?
    \item[\textbf{RQ2:}] How does listeners' use of motivational interviewing techniques change with their counseling experience?
\end{enumerate}

To answer these questions, we have annotated a corpus of 734 conversations comprising  14,797 messages (utterances) with MI codes using a multi-label approach. Additionally, we  defined new classes to capture other language behavior common in OMHCs. Manually labeling millions of utterances is a laborious task; therefore, we built  machine learning classifiers to measure the MI and other language behaviors at scale. This paper uses  7Cups data to build mental-health domain specific models. We then use the annotated corpus data to fine-tune these domain specific, transformer-based large language models \cite{intro_all_you_need_is_attention} to predict the MI code of any given listener utterance. This allows us to perform our analysis at a scale of millions of conversations and utterances.

For RQ1, we first grouped conversations on 7Cups  into two categories: "\emph{satisfactory}" ones with rating of 4 or 5 and "\emph{unsatisfactory}" ones with a rating of 1 to 3. Then we labeled each conversations' utterances using the fine-tuned classifiers and use logistic regression analysis to determine which MI techniques best predict whether conversations were satisfactory or unsatisfactory. 
RQ2, along with results from RQ1, tests whether listeners improve their behavior as they gain more experience. We analyzed this by identifying users who have been active for over a year. We then pinpoint which MI techniques become more/less prevalent as a listener gained experience. 

While there is prior Human-Computer Interaction (HCI) research focusing on the identification of various strategies for online counseling success \cite{intro_prev_online_mh_papers, intro_prev_online_mh_papers_silver_cloud}, our work extends this by measuring both MI techniques frequently used offline and  online-specific supportive strategies to predict conversational satisfaction. 
Results from millions of online conversations can  extend the existing literature on the use of motivational interviewing by professional therapists to understand what works online. 

\section{Related Work}
\subsection{Peer-to-Peer Counseling in Online Communities} 
Research has shown that peer support in OMHCs can lead to increased feelings of belonging, connectedness, and insight \cite{fut_mh_care}. Additionally, peer-support can offer a solution to the historical inaccessibility of traditional mental health services, which has resulted in a majority of mental problems being untreated \cite{mental_disorders_US}. Prior research on peer-to-peer counseling has shown the efficacy of peer-counseling for clients. For example, recent work on 7Cups compared users' satisfaction with peer counseling to more formal psychotherapy and found that users' perceived peer-counseling to be as effective as psychotherapy \cite{intro_online_support_works}. Users also reported that they felt peer-counselors were more genuine than professional psychotherapists \cite{intro_online_support_works}. \citet{race_OMHC} has additionally shown the importance of peer-counseling for marginalized communities, such as the LGBTQ+ community and racial minorities.

The value of peer-counseling in OMHCs , however, probably depends upon the training and skill level of each peer-counselor. Under-trained volunteer counselors could lead to harmful sessions, as research has found that therapy strategies such as rigidity and over-control led to clients reporting they felt disempowered and devalued \cite{therapy_harm}. As such, there is a growing need for peer-counseling training programs on OMHCs. In offline settings, research conducted on peer-counselor training within a senior community reported a statistically significant improvement in peers' knowledge and skills following training \cite{senior_training_program}. While peer-counseling OMHCs vary in the degree and presence of listener training, some OHMCs such as 7Cups have introduced training programs for listeners based in active listening (an approach where counselors utilize open-ended questions to guide the conversation) and motivational interviewing \cite{7cups, YAO_CMU}. However, previous research on OMHC training programs found that volunteers on 7Cups reported feeling unprepared \cite{YAO_CMU} and online volunteer listeners for OMHCs are typically not as well trained as volunteer listeners who conduct peer-counseling offline \cite{moments_of_change}. These discrepancies indicate the need for improved training programs that are tailored to the specific online environment where volunteer listeners will be working. 

\subsection{Identifying Techniques for Online Counseling Success}
Due to the above described popularity of OMHCs, recent HCI research has focused on what counseling techniques lead to success \cite{Saha_Sharma_2020, mh_tenure, intro_prev_online_mh_papers, intro_prev_online_mh_papers_silver_cloud}. The counseling success of OMHCs has been defined in terms of a variety of metrics, from user churn \cite{user_churn} and retention \cite{mh_tenure} to more complex measures of affective, behavioral, and cognitive psychosocial change \cite{Saha_Sharma_2020}.
Due to the complexity and nuance of measuring success within online counseling sessions, we focus on client satisfaction as our measure of success, in the tradition of prior work \cite{intro_prev_online_mh_papers, client_satis_1, intro_online_support_works}. Furthermore, this self-reported metric is linked to the degree of therapeutic alliance, which is a standard metric in this area and associated with client retention \cite{ther_alliance, martin_ther_alliance, sharf_ther_alliance}. Client satisfaction differs from therapeutic alliance, which specifically measures the relationship between support seeker and listener, whereas client satisfaction is only one of many potential indicators of therapeutic alliance. 
It is also important to note that even therapeutic alliance has certain shortcomings in measuring the overall "success" of therapeutic sessions; while therapeutic alliance is predictive of user engagement, retention, and behavior change during counseling, it is not clear whether therapeutic alliance is a predictor of clients sustaining behavioral changes after counseling \cite{meier2005role}.

Saha et al. \cite{Saha_Sharma_2020} found the primary predictors of success are complex language factors such as adaptability and style. Similarly, consistent communication \cite{mh_tenure}, linguistic style-matching \cite{intro_prev_online_mh_papers}, and concrete, positive, supportive feedback from listeners \cite{intro_prev_online_mh_papers_silver_cloud} are all reported indicators of successful counseling sessions. While past research focused on detecting patterns correlated to counseling success on OMHCs, not much attention has been paid to analyzing counseling sessions with a comprehensive framework, with a few exceptions like \cite{atkins, perez-rosas-etal-2018-analyzing, perez-rosas-etal-2019-makes}.
Our work extends this line to work by analyzing session satisfaction at scale through a predefined MI schema.

\subsubsection{Client-Centered Counseling} Our research is based in a specific form of client-centered counseling that adds a directional component called motivational interviewing \cite{MI_original}. Client-centered counseling is a counseling approach developed by Carl Rogers that emphasizes active listening and a non-directive conversational style \cite{rogers_cct}. Client-centered strategies are effective in treating a variety of mental health problems such as addiction, anxiety, and depression \cite{person_centered_evaluation} and are the current chosen approach for OMHCs such as 7Cups \cite{7cups}. The quality of client-centered MI counseling sessions has been explored in offline counseling media through language modeling and fine-grained analyses of MI codes \cite{perez-rosas-etal-2018-analyzing, perez-rosas-etal-2019-makes, person_centered_evaluation}. We extend from these fine-grained analyses to analyze how MI strategies apply to OMHCs. 

Harrison and Wright conducted one such study on client-centered counselors within a text-based online environment and found counselors developed stylistic changes distinct from their in-person approaches, indicating there may be significant differences in client-centered strategies used online versus offline. Additionally, \cite{video_conferencing_person_centered} conducted an analysis of client-centered counseling over video conferencing and reported counselors struggled more with establishing connections with clients and noted the increased difficulties for clients to perceive acceptance and empathy from their counselors. Despite these difficulties,  \cite{video_conferencing_person_centered} ultimately concluded that the condition for successful client-centered therapy can be met effectively in online media, pointing to the potential for client-centered therapy to be utilized effectively on peer-counseling platforms. To further explore the usage of client-centered counseling online and provides insights into how the challenges of online counseling may be effectively met through MI. 


\subsubsection{AI for Online Mental Health Communities} Researchers are starting to build integrated artificial intelligence (AI) application to improve the effectiveness of those participating in online support platforms \cite{MepsBot, AI_MOST, MI_flagging}. 
Many technological aids to assist in online writings like MepsBot have been implemented in different domains \cite{AI_metaphor_creation, AI_mirrorU, AI_social_media} and have potential within OMHCs to aid volunteer counselors to improve sessions. 
Our research aims to offer insights that may potentially be incorporated into a similar AI writing assistant.
While the direct incorporation of AI applications into OMHCs is still in preliminary stages, there have been several applications of AI to analyze linguistic patterns and explore factors that might be responsible for counseling success \cite{intro_prev_online_mh_papers, linguistic_accommodation}. 

AI has also been specifically applied to the domain of understanding and improving MI counseling offline. Atkins et al. \cite{atkins} utilized topic modeling to predict MI codes. Imel et al. \cite{imel} explored the feasibility of an automated, machine-learning based feedback system for MI, and Flemotomos et al. \cite{Flemotomos2021AmIA} similarly explored automatically annotating MI codes to predict a therapist's performance both at the utterance and session levels. Our work continues this line further to explore text-based online therapy (versus dialogue) and create a new MI framework specific to online counseling. Our research builds on prior work focusing  on automatic MI annotation and its applications  and attempts  to improve MI model performances and additionally answers several RQs that explore the nuances of online counseling.  


\section{Motivational Interviewing Treatment Integrity (MITI) Framework}
Our research is based on a client-centered counseling style called \textit{motivational interviewing} \cite{MI_original}. Motivational interviewing has long been an effective treatment for substance abuse and addiction \cite{MI_original}. More recently, motivational interviewing has become a popular schema for peer-to-peer counseling in platforms such as 7Cups \cite{7cups}. Given the effectiveness of motivational interviewing and its use in OMHCs, the goals of this paper are to measure MI skills at scale, assess which MI skills predict successful client evaluation of counseling sessions, and examine how use of these skills evolve over listeners' tenure. To do so, we use motivational interviewing skill codes, which are a set of predefined behavioral codes, to categorize listener utterances in a conversation \cite{MISC}. 
To meet the unique requirements of an online peer-counseling platform, we create a MITI Framework of thirteen MI codes aggregated from three motivational interviewing manuals \cite{MITI, MISC, MITI_4} and supplement them with four newly created codes discussed in Section~\ref{sec:dataset_categories}. The MITI Framework is described and summarized in Table \ref{tab:MITI_framework}. The examples in the table are taken from conversations on 7Cups and paraphrased using round-trip translation to ensure anonymity. 

Our framework follows the traditional MISC guidelines and divides the codes into three categories: (1) \textsc{MI-consistent} ones that promote the client-centered nature and goal of MI; (2) \textsc{MI-inconsistent} ones  that undermine this goal either through inappropriate behavior or by explicitly telling the client what to do without collaborating or emphasizing the client's autonomy; and (3) \textsc{Others} that serve primarily to establish rapport with the client during a session \cite{MITI_change_talk}. In addition to categories defined in Table \ref{tab:MITI_framework}, we initially also included 4 other codes detailed in MI manuals - \emph{Structure}, and the MI-inconsistent codes: \emph{Warn}, \emph{Confront} and \emph{Raise Concern}. However, these classes rarely occurred in the 7Cups conversations we analyzed, with fewer than 50 per class among 14,797 listener utterances. With so few cases, they could not be modeled in the classification task and were therefore removed from our framework. 

\begin{table}[]\centering
\resizebox{1\columnwidth}{!}{%
\begin{tabular}{|lll|}
\hline
\multicolumn{1}{|l|}{\textbf{Code}}                                                       & \multicolumn{1}{l|}{\textbf{Description}}                                                                                                                                                                          & \textbf{Examples}                                                                                                                                                       \\ \hline
\multicolumn{3}{|l|}{\textbf{MI-Consistent}}                                                                                                                                                                                                                                                                                                                                                                                                                                             \\ \hline
\multicolumn{1}{|l|}{Affirm}                                                              & \multicolumn{1}{l|}{\begin{tabular}[c]{@{}l@{}}The listener says something positive or \\ complimentary. It may serve as reinforcement.\end{tabular}}                                                              & \emph{\begin{tabular}[c]{@{}l@{}}This is such a big step forward! \\ I am very proud of you.\end{tabular}}                                                                                                                                       \\ \hline
\multicolumn{1}{|l|}{\begin{tabular}[c]{@{}l@{}}Emphasizing \\ Autonomy\end{tabular}}     & \multicolumn{1}{l|}{\begin{tabular}[c]{@{}l@{}}The listener focuses responsibility on \\ the member.\end{tabular}}                                                                                            & \emph{\begin{tabular}[c]{@{}l@{}}OK, this is your call.\end{tabular}}               \\ \hline
\multicolumn{1}{|l|}{\begin{tabular}[c]{@{}l@{}}Open\\ Question\end{tabular}}             & \multicolumn{1}{l|}{\begin{tabular}[c]{@{}l@{}}Questions that are not closed and leave \\ a latitude for response.\end{tabular}}                                                                                   & \emph{What brings you here today?}                                                                                                                                \\ \hline
\multicolumn{1}{|l|}{\begin{tabular}[c]{@{}l@{}}Closed \\ Question\end{tabular}}          & \multicolumn{1}{l|}{A question that implies a short answer.}                                                                                                                                                       & \emph{\begin{tabular}[c]{@{}l@{}}Have you sought professional help \\ with these triggers?\end{tabular}}                                                                       \\ \hline
\multicolumn{1}{|l|}{\begin{tabular}[c]{@{}l@{}}Persuade (with \\ Permission)\end{tabular}} & \multicolumn{1}{l|}{\begin{tabular}[c]{@{}l@{}}Listener makes overt attempt to change member's\\ opinions, attitudes, or behavior. Listener asks for\\  permission first or emphasizes collaboration.\end{tabular}} & \emph{\begin{tabular}[c]{@{}l@{}}Unfortunately, I am not a doctor, \\ but it is always best to get a second opinion \\ from someone who is at least believable.\end{tabular}}                                                        \\ \hline
\multicolumn{1}{|l|}{Reflection}                                                          & \multicolumn{1}{l|}{\begin{tabular}[c]{@{}l@{}}Reflections capture and return to the member\\ something the member has said.\end{tabular}}                                                                         & \emph{\begin{tabular}[c]{@{}l@{}}You feel dysphoric and \\ it causes you suffering.\end{tabular}}                                                                       \\ \hline
\multicolumn{1}{|l|}{\begin{tabular}[c]{@{}l@{}}Seeking \\ Collaboration\end{tabular}}    & \multicolumn{1}{l|}{\begin{tabular}[c]{@{}l@{}}Listener explicitly attempts to share power or \\ acknowledge the expertise of the member.\end{tabular}}                                                            & \emph{\begin{tabular}[c]{@{}l@{}}May I ask what decisions \\do you feel bad about?\end{tabular}} \\ \hline
\multicolumn{3}{|l|}{\textbf{MI-Inconsistent}}                                                                                                                                                                                                                                                                                                                                                                                                                                           \\ \hline
\multicolumn{1}{|l|}{Direct}                                                              & \multicolumn{1}{l|}{Counselor gives an order or command.}                                                                                                                                                          & \emph{\begin{tabular}[c]{@{}l@{}}You need to go somewhere else \\and get a second opinion.\end{tabular}}                                                                                                                                  \\ \hline
\multicolumn{1}{|l|}{Inappropriate*}                                                      & \multicolumn{1}{l|}{\begin{tabular}[c]{@{}l@{}}Excessive swear words, asking unnecessary \\ personal information, abusive behavior, etc.\end{tabular}}                                                 & \emph{What are you wearing?}                                                                                                                                              \\ \hline
\multicolumn{3}{|l|}{\textbf{Other}}                                                                                                                                                                                                                                                                                                                                                                                                                                                     \\ \hline
\multicolumn{1}{|l|}{Grounding}                                                           & \multicolumn{1}{l|}{Facilitate or acknowledge.}                                                                                                                                                                    & \emph{Okay or Sure or Hmm}                                                                                                                                                  \\ \hline
\multicolumn{1}{|l|}{\begin{tabular}[c]{@{}l@{}}Giving\\ Information\end{tabular}}             & \multicolumn{1}{l|}{\begin{tabular}[c]{@{}l@{}}Educate, provide feedback, \\or expresses a professional opinion \\
without persuading, advising, or warning.\end{tabular}}

& \begin{tabular}[c]{@{}l@{}}\emph{There are some great meditation links on here.}\end{tabular}                                                                                                                                \\ \hline

\multicolumn{1}{|l|}{Support*}                                                            & \multicolumn{1}{l|}{\begin{tabular}[c]{@{}l@{}}Sympathetic, compassionate, or \\understanding comments\end{tabular}}                                                                                                                                         & \emph{I'm always here to help you.}                                                                                                                                              \\ \hline
\multicolumn{1}{|l|}{\begin{tabular}[c]{@{}l@{}}Personal\\ Disclosure\end{tabular}}       & \multicolumn{1}{l|}{\begin{tabular}[c]{@{}l@{}}When the listener shares something personal as \\part of  the conversation.\end{tabular}}                                                                               & \emph{\begin{tabular}[c]{@{}l@{}}I know exactly what you mean,\\ I was like that a few years back.\end{tabular}}                                           \\ \hline
\multicolumn{1}{|l|}{Introduction}                                                        & \multicolumn{1}{l|}{Greetings, exchanging names, etc.}                                                                                                                                                             & \emph{Hello welcome to 7COT. My name's John}                                                                                                                                     \\ \hline
\multicolumn{1}{|l|}{Conclusion}                                                          & \multicolumn{1}{l|}{Statement indicating that the listener need to leave.}                                                                                                                                        & \emph{I'm ending our chat now.}                                                                                                                                        \\ \hline
\multicolumn{1}{|l|}{Chit-Chat*}                                                           & \multicolumn{1}{l|}{\begin{tabular}[c]{@{}l@{}}Connecting with members by conversing about\\ topics not related to the general session.\end{tabular}}                                                              & \emph{I wonder if we can fly one day.}                                                                                                                                              \\ \hline
\multicolumn{1}{|l|}{Other}                                                               & \multicolumn{1}{l|}{Any utterance that does not fit into the above.}                                                                                                                                       &                                                                                                                                                                         \\ \hline
\end{tabular}
}
\caption{MITI Framework; *denotes newly created codes.}\label{tab:MITI_framework}
\end{table}

\section{dataset}
\label{sec:dataset}
7 Cups of Tea is an online platform that provides free counseling to those experiencing emotional distress \cite{7cups} by connecting volunteer counselors to support seekers on an anonymous chatroom for counseling. The listeners on 7Cups are volunteers who receive 30-60 minutes of training and must pass an exam, but are not licensed professionals. This structure allows 7Cups to offer free, accessible counseling, but with minimally trained volunteers who may not be as equipped as professionals to handle members dealing with emotional distress and complex mental health problems. After each conversation, members have the option to rate their conversation with the listener on a 5-point Likert scale from 1 (worst) to 5 (excellent) to indicate how satisfied they were with the conversation.

This work obtains the chat messages sampled from the 7Cups platform for the annotation of the dataset, via a research collaboration with 7 Cups of Tea. The annotated dataset consists of 14,797 labeled unique listener utterances (or 734 conversations) from the platform. These listener utterances are labeled according to the motivational interviewing framework described in Table \ref{tab:MITI_framework}. 
The distribution of the MI techniques and their sizes are given in Table \ref{tab:class_description}. Since we utilized a multi-label approach, in which each utterance may receive up to three MI code labels, the distribution counts  in Table \ref{tab:class_description} sum to more than the 14,797 unique utterances labeled.  

\begin{table*}
\centering
  \begin{tabular}{lcc}
    \toprule
    Category & \#Utterances & Annotation Agreement\\
    & & (Krippendorff’s alpha)\\
    \midrule
    Giving Information & 304 & 1\\
    Reflection & 1697 & 0.585\\
    Affirm & 346 & 0.659\\
    Closed Question	& 1956 & 0.868\\
    Open Question & 2507 & 0.950\\
    Persuade & 1918 & 0.424\\
    Direct  & 250 & 0.666\\
    Emphasizing Autonomy  & 120 & 0.665\\
    Grounding & 1027 & 0.826\\
    Personal Disclosure & 1605 & 0.658\\
    Introduction/Greeting & 1260 & 0.877\\
    Conclusion & 340 & 0.791\\
    Seeking Collaboration/Permission & 271 & 0.662\\ \hline
    Support & 1493 & 0.842\\
    Inappropriate & 224 & 1\\
    Chit-Chat & 963 & 0.59\\
    Other & 1299 & 0.664\\
    \bottomrule
  \end{tabular}
  \caption{Distribution of categories}   \label{tab:class_description}
\end{table*}

\subsection{Motivational Interviewing Categories}
\label{sec:dataset_categories}
Listener utterances are annotated using 17 class categories as presented in Table \ref{tab:class_description}. Note that the distribution of classes is highly imbalanced. Some of the classes like \emph{Open Questions},  \emph{Closed Questions} and \emph{Personal Disclosure} have more than 1500 datapoints each, whereas classes like \emph{Emphasizing Autonomy}, \emph{Inappropriate}, and \emph{Direct} have less than 300 data points. This class imbalance is characteristic of client-therapist conversations and previous works have shown a similar trend \cite{da_therapy_conversation}. This class skewness presents a challenge for developing listener behavior in natural language processing models. We use the one versus all binary classification approach described in Section \ref{sec:labeling} to deal with this class skewness during automated labeling.
In addition to the motivational interviewing codes obtained from the MITI Manual \cite{MITI}, we have created 4 new codes to more accurately describe the listener utterances in this online context: \emph{Support},  \emph{Inappropriate}, \emph{Chit-Chat}, and \emph{Other}, 
and define them as follows:
\begin{itemize}
    \item \textbf{Support:} This class describes utterances that are generally sympathetic, compassionate, or understanding comments. Utterances following the \emph{Support} technique are characterized by "\emph{agreeing with the client}". Few examples of \emph{Support} are: "\emph{You've got a point there}" (agreement), "\emph{That must have been difficult}" (compassion), "\emph{I can see why you feel this way}" (understanding).
    \item \textbf{Inappropriate:} This class refers to utterances that include swear words, asking unnecessary personal identifiable information, or abusive behavior. The creation of this class was especially motivated by 7Cups being an online, anonymous chat platform, which often includes inappropriate comments. Therefore, identifying these comments is an important step in creating a better environment for both clients and listeners. 
    \item \textbf{Chit-Chat:} This class was also motivated by the unique medium of an online chat. \emph{Chit-Chat} is a non-traditional therapeutic code that describes utterances where jokes are made or the listener otherwise attempts to connect to the client in a manner that is not directly related to mental health or the topic at hand. Although this strategy is not traditionally therapeutic, \emph{Chit-Chat} may serve to make the client more comfortable with the listener and could potentially have positive affects on the conversation's success, which is a topic we intend to explore further. 
    \item \textbf{Other:} This class serves as a label to capture utterances that do not fit accurately into the other 16 categories. Example utterances may be typos or short responses such as, "Good" that are not neutral enough to be labeled \emph{Grounding}.
\end{itemize}

\subsection{The Annotation Process}
Building the corpus required several rounds of annotation. The initial annotation process had three components. First, three annotators individually labeled a series of conversations. Second, we calculated Krippendorff's alpha \cite{krippendorff2011computing} for each label to measure agreement between annotators. We considered a Krippendorff's alphas of approximately 0.7 as the cutoff of a reasonable level of agreement for a specific MI label, building on the rule of thumb of the Krippendorff's alpha \cite{krippendorff2011computing} score. Lastly, the annotators went through disagreements, clarified misconceptions, and decided on the correct label(s) for an utterance where there had been disagreement. 

Once the Krippendorff's alphas reached the threshold of  0.7 for most of the classes (Table \ref{tab:class_description}), we determined that there was a consistent, accurate understanding of the MI codes and there was enough agreement between the annotators to begin labeling separately. This allowed us to significantly speed-up the process of labeling utterances. Additionally, to further increase the rate of labeling, we supplemented manual labeling with a semi-supervised approach using machine learning classifiers based on a large-pretrained language model called BERT \cite{BERT}. BERT is a neural network architecture that outperforms other architectures on a variety of downstream language processing tasks, such as classification, due to its ability to capture context specific semantics of language. BERT and BERT-like models capture task agnostic semantics of words due to the  unsupervised pre-training on large amounts of textual information like Wikipedia articles. To speed up the annotation task, we further fine-tuned the BERT model to capture mental health specific word
usage and the labeling process. For each batch of conversations, the listener utterances were inputted into each BERT classifier. We defined a threshold of 0.7 confidence and only looked at the utterances the classifiers predicted as belonging to a certain class with $>=$ 0.7 confidence. This cutoff was chosen based on the annotator's qualitative analysis of the classifiers' outputted labels; it was observed that if a model labeled an utterance with $<$0.7 confidence, it was rarely, if ever, correct when manually verified. The annotators then manually verified each of the classifiers' outputs to ensure their correctness before assigning a label to the utterance. We found that this semi-supervised process was very effective for the labels: \emph{Personal Disclosure}, \emph{Grounding}, \emph{Introduction}, \emph{Closed Question}, and \emph{Open Question}, most likely because these class rely less heavily on the context of previous chat messages than others. The remaining techniques were primarily labeled manually. This combination of manual and semi-supervised annotation allowed us to increase the corpus from 1,539 labeled listener utterances (or 147 conversations) to 14,797 labeled listener utterances (or 734 conversations).

\section{Automated labeling using Large Language Models} We implement automated labeling using large scale language models fine-tuned on the data described in Section \ref{sec:dataset}. This allows us to analyze our research questions at the scale of millions of utterances. Constructing this automation pipeline required several considerations, described in the following paragraphs, to meet the nuances of an OHMC. 
\label{sec:labeling}

\begin{table*}[h]

  \begin{tabular}{p{3.5cm}p{3cm}p{3cm}p{3cm}}
     \toprule
    Model Name & No previous context & 1 previous utterance as context & 5 previous utterances as context \\
    \midrule
    BERT-Base-Uncased & 0.570 & 0.600 & 0.539 \\
    BERT-large & 0.605 & 0.639 & 0.608 \\
    RoBERTa-Base & 0.608 & 0.628 & 0.606 \\
    BERTweet & 0.642 & 0.693 & 0.649 \\
    \textbf{Mental Health BERT} & \textbf{0.678} & \textbf{\textit{0.725}} & \textbf{0.705} \\
    \bottomrule
  \end{tabular}
  \caption{Averaged F1 Scores across MI Codes for Different Models without Domain Specific Pretraining}
  \label{tab:model_averages}
\end{table*}


\begin{table*}[h]

\resizebox{1\columnwidth}{!}
{
  \begin{tabular}{p{3.5cm}p{1.6cm}p{1.5cm}p{1.5cm}{l}}
     \toprule
    MI Code & No context & 1 previous utterance & 5 previous utterances  & Most Relevant Words\\
    \midrule
    Giving Information & \textbf{0.656} & 0.574 & 0.387 & mean, tell, ok, aww, oh\\
    Reflection  & 0.631 & 0.692 & \textbf{0.780} & like, yeah, know, think, feel\\
    Support & 0.721 & 0.750 & \textbf{0.778} & understand, okay, hope, feel, sorry\\
    Affirm & \textbf{0.677} & 0.624 & 0.392 & good, great, nice, awesome, love\\
    Closed Question & 0.905 & \textbf{0.915} & 0.895 & ok, talk, like, know, want\\
    Persuade & 0.661 & 0.770 & \textbf{0.841} & like, think, know, happen, thing \\
    Open Question & 0.918 & \textbf{0.943} & 0.938 & like, happen, tell, mean, think \\
    Seeking Collaboration & \textbf{0.518} & 0.500 & NA & ask, tell, maybe, mind, want\\
    Inappropriate & 0.566 & 0.667 & \textbf{0.783} & kiss, big, eat, wet, wanna\\
    Direct & 0.516 & \textbf{0.523} & 0.514 & need, dont, stop, talk, tell\\
    Emphasizing Autonomy & 0.730 & \textbf{0.785} & NA & choice, want, best, decide, right\\
    Grounding & 0.773 & \textbf{0.846} & 0.835 & ok, yeah, oh, cool, hmmm\\
    Personal Disclosure & 0.803 & 0.776 & \textbf{0.817} & good, know, want, yeah, time\\
    Introduction & 0.899 & \textbf{0.927} & 0.851 & hello, good, morning, today, welcome\\
    Conclusion & \textbf{0.806} & 0.756& 0.500 & sleep, bye, care, gonna, later\\
    Chit-Chat & 0.307 & \textbf{0.652} & 0.497 & NA \\
    Other & 0.458 & 0.633 & \textbf{0.770} & NA\\
    \bottomrule
    Averaged F1  & 0.678 & \textbf{0.725} & 0.705 & NA\\
    \bottomrule
  \end{tabular}
}
\caption{Best Model Metrics (MH-BERT): F1 scores of the Positive Classes for MI Codes using 0, 1, and 5 utterances as previous context when using \textit{domain specific} Mental Health BERT}
\label{tab:best_model_mh}
\end{table*}
Language models learn language patterns from the data using statistical  methods. Such models are useful for many machine learning prediction tasks. Recently, large scale language models trained on unlabeled textual data have achieved impressive results on classification tasks \cite{large_language_models, BERT, roberta}. In this study, we train a classifier using these large language models and fine-tune it using our annotated dataset. This allows us to automatically label listener utterances with MI codes. 


We use several state--of-the-art architectures of large-scale language models from the Huggingface library \cite{huggingface}: BERT-base-uncased (base-line), BERT-large-uncased \cite{BERT}, RoBERTa-base \cite{roberta}, Vinai/BERTweet \cite{bertweet}. Our experiments compare the models (Appendix Tables \ref{tab:f1scores_1}, \ref{tab:f1scores}) and show that Vinai/BERTweet architecture has the best F1 score as it is pre-trained on Twitter data which matches the informal short text messaging format of the 7Cups platform. To capture domain specificity, we build a domain specific instance (known as Mental Health BERT) by further pre-training the Vinai/BERTweet checkpoint (best performing model) on 120 million chat messages from 7Cups. 

For training the classifier, we build a classifier for each MI code using a one versus all method to deal with the large number of classes. Thus, an utterance can have multiple MI codes to address the issue of low representation of classes in our dataset. An example utterance for multiple MI codes would be: "\emph{Hi, how are you doing today?}", which has two MI codes, \emph{Introduction} and \emph{Open Question}. 
For each of the 17 MI codes, we measure the F1 score of the positive class (the presence of an MI technique in an utterance) for evaluation. The metrics of accuracy, precision, and recall are not best suited for highly imbalanced data, especially for a binary classification task such as ours since models tend to overfit towards the class with greater representation. 

Human annotation  for some MI codes, such as \emph{Support}, \emph{Reflection}, and \emph{Giving Information}, required previous context to determine the correct micro-skill for an utterance. To examine the role of context in developing accurate machine learning models, we fine-tune them by adding the previous context in our data by looking at 0 (no previous context), 1, and 5 previous utterances. The addition of context is done by concatenating the previous \textit{k} utterances to the current utterance (where \textit{k} belongs to \{0, 1, 5\}). The context is independent of the person who wrote the utterance (i.e., the member or the listener). 
Using this setup, Table \ref{tab:model_averages} gives the averaged F1 scores of the respective MI codes as positive classes. The F1 scores of MI codes as positive classes for the best model are shown in Table \ref{tab:best_model_mh}. The F1 scores of MI codes for other models are given in the Appendix (Table \ref{tab:f1scores_1},  \ref{tab:f1scores}). We highlight the key takeaways of the model development process:
\begin{itemize}
    \item Best Architecture - Mental-Health-BERT (based on vinai/bertweet \cite{bertweet}, Table \ref{tab:best_model_mh}): The MH-BERT model works substantially better than the models not trained on the domain specific data, as observed in Table \ref{tab:model_averages}. 
    \item Optimum Previous Messages as Context: Table \ref{tab:model_averages} shows that using 1 previous utterance for additional context has the best classification performance, regardless of the choice of model architecture.
    \item Inappropriate: For \textit{Inappropriate}, we finetune HateBERT \cite{hatebert} for few shot classification, along with other models, as HateBERT is trained on abusive language that is fairly similar to the inappropriate text found in the 7Cups data. HateBERT gives an F1 score for the positive class of 0.6061 (No context), 0.7119 (1 previous utterance as context), 0.8421 (5 previous utterances as context). These metrics are much better than the other models discussed above.
    \item Presence of NA Values: Some of the cells in Table \ref{tab:best_model_mh} contain NA values because there is not enough representation of the utterances in the annotated data for the particular MI technique to have 5 previous utterances. For example,\textit{ Seeking Collaboration} has only 27 positive data points when considering 5 previous utterances as context; this makes it very difficult to build robust classifiers due to limited training data. 
\end{itemize}

As we use automated labeling without manual verification to explore our research questions, we seek to ensure that the quality of the labeling is maintained when automated. Therefore we randomly sampled 100 annotated utterances and manually analyzed the results. Two authors annotate the hundred samples, we calculate inter-rater agreement and the agreement with the model for validation. After comparing model labeling with human analysis, we find a Krippendorff's Alpha agreement score of 0.614 (Table \ref{tab:annotation_agreement_validation}). We observe perfect inter-rater agreement for certain MI codes due to their low representation in the data sample. As a sanity check, we also retrieve the top words of each MI code according to the term frequency-inverse document frequency scores. For this step, we removed stop words and lemmatized all the words for standardization of verb forms. The top words for each MI code are given in Table \ref{tab:best_model_mh}. The words generated by the natural language processing models are consistent with the definitions of the codes (Table \ref{tab:MITI_framework}), giving distinct but relevant top words for at least 10 out of the 15 codes under consideration (excluding \emph{Chit-Chat} and \emph{Other}). This is true for codes like \emph{Introduction}, \emph{Conclusion}, \emph{Emphasizing Autonomy}, \emph{Affirm}, \emph{Direct}, \emph{Open Questions}, \emph{Closed Questions}, \emph{Support}, \emph{Inappropriate} and \emph{Grounding}. We do not show words for \emph{Chit-Chat} and \emph{Other}, as these two MI techniques are meant to encapsulate utterances that are diverse in nature in terms of the topics discussed. These robustness checks confirm the efficacy of our automated pipeline. 


\section{Analyses}
For our analyses, we utilize the Mental Health BERT model described above to predict MI codes on a sampled set of conversations from 7Cups taken over a six-month time-span (January 2020 to August 2020). Since we are interested in understanding what counseling strategies in a conversation leads to client satisfaction, we convert satisfaction ratings to two categories: \emph{satisfactory} conversations (received a rating or 4 or 5) and \emph{unsatisfactory} conversations (received a rating of 1-3). The threshold of 4 for a satisfactory conversation was chosen based on the average rating across all conversations in our dataset being 4.14.
\subsection{RQ1: What Motivational Interviewing Techniques Predict More Satisfaction?}
\label{sec:RQ1}
To answer RQ1, we study the association between listener MI usage and satisfying conversations. Conversations are categorized into satisfactory  or unsatisfactory ones, as displayed in Table \ref{tab:RQ1 buckets}. The average rating for these conversations is 4.14, indicating that members may be more likely to leave a rating if the conversation was satisfactory. In total, we analyze 63,012 conversations (5,632,402 utterances) to explore what MI techniques lead to conversation success. 
\subsubsection{Logistic Regression Analysis of MI Codes and Satisfactory Conversations} After obtaining MI label(s) for each utterance in each conversation, we run a logistic regression model to predict the outcome of a conversation (1 = satisfactory conversation, 0 = unsatisfactory conversation). We use the number of times each MI code occurs in a given conversation as our independent variables. To account for the subjectivity of using a rating scale from 1-5 (as certain members may be more likely to rate all conversations lower or higher), we use member's past average rating as a control variable. We also use listener age and member age as control variables and find in Table \ref{tab:RQ1_LR_Results} that every control variable used is statistically significant, with member's average past rating having an especially high odds ratio of 1.685, indicating the importance of controlling for a member's rating tendency. To account for the imbalanced counts of our sample data listed in Table \ref{tab:RQ1 buckets}, the classes in our logistic regression model are weighted accordingly, using the inverse of the respective class frequencies. 
\begin{table}[h]

\begin{tabular}{p{4cm} p{2cm} p{3.2cm}}
 \hline
 Conversation Category & Count & Percent of Total Count \\ 
 \hline
 Satisfactory & 49,892 & 79\% \\ 
 Unsatisfactory & 13,120 & 21\%   \\
 \hline
\end{tabular}
\caption{ Breakdown of samples used per conversation category for analysis in Section \ref{sec:RQ1}.}\label{tab:RQ1 buckets}
\end{table}


 
\begin{table}[h]

\begin{tabular}{p{6cm} p{2cm} p{2cm}}
 \hline
 \textbf{MI Code} & \textbf{Coefficient} & \textbf{Odds Ratio} \\ 
 \hline
 \textbf{MI-Consistent} &  \\
 \hline
 Affirm &  0.036* & 1.037  \\ 
 Emphasizing Autonomy & 0.078 $\odot$ & 1.081   \\
 Open Question &  0 & 1 \\ 
 Closed Question & 0.011 $\odot$ & 1.012\\ 
 Persuade (with permission) & 0.018** & 1.019   \\ 
 Reflection & \textbf{0.035***} & \textbf{1.035}\\ 
 Seeking Collaboration & 0 & 0.999 \\
 \hline
 \textbf{MI-Inconsistent} & \\
 \hline
 Direct &  0.019 &  1.019 \\
 Inappropriate & \textbf{-0.086***} & \textbf{0.917} \\ 
 \hline
 \textbf{Other} & \\
 \hline
 Grounding & 0.008 $\odot$ & 1.008 \\ 
 Giving Information & -0.025 & 0.975   \\
 Support & 0.013 & 1.013 \\ 
 Personal Disclosure & 0.001  & 1.001 \\
 Introduction & 0 & 1 \\
 Conclusion &  0.014 & 1.015 \\ 
 Chit-Chat & -0.004 & 0.996 \\ 
 \hline
 \textbf{Control Variables} \\
 \hline
 Member Age & -0.007*** & 0.993 \\
 Listener Age & -0.003*** & 0.997  \\
 Member Past Average Rating & 0.522*** & 1.685 \\
 \hline
 \hline
 AIC of Model & 97,800 \\
 \hline
 
\end{tabular}
\caption{\label{tab:RQ1_LR_Results} Associations between strategies and satisfactory conversations. ***p < 0.001; **p<0.01; *p<0.05 $\odot$p<0.1. 
}
\end{table}

Table \ref{tab:RQ1_LR_Results} shows the coefficients and odds ratio of each motivational interviewing technique resulting from the logistic regression model. We observe that \emph{Reflection}, \emph{Persuade}, and \emph{Affirm} are statistically significant predictors for client satisfaction while \emph{Inappropriate} is a statistically significant negative predictor for client satisfaction. \emph{Reflection} is the most statistically significant positive predictor and has an odds ratio of 1.03, which indicates that each additional usage of \emph{Reflection} in a conversation (a one unit increase) increases the odds of the conversations receiving a satisfactory rating by $3\%$. It is important to note that each of the significant positive predictors for client satisfaction (\emph{Reflection}, \emph{Persuade}, \emph{Affirm}) are MI-consistent, aligning with previous research on the success of MI-consistent codes in standard offline counseling sessions \cite{magill_MI_causal_model}. 

In prior work, success within MI counseling is measured by the prevalence of "change talk" defined by Miller and Rolnick as, "any self-expressed language that is an argument for change" and a lack of "sustain talk" defined as, "the person’s own arguments for not changing, for sustaining the status quo" \cite{MI_original}. A previous study on which MI codes were most likely to be followed by "change talk" and/or "sustain talk" found that \emph{Reflection} was significantly more likely to be followed by client "change talk", aligning with our finding that \emph{Reflection} is a significant indicator of client satisfaction \cite{MITI_change_talk}. However, this same research found that "sustain talk" was also more likely to follow usage of \emph{Reflection}, which may indicate that while \emph{Reflection} leads to client satisfaction, there are issues with this MI strategy in regards to its relationship with "sustain talk". The dual nature of \emph{Reflection} positively being followed by "change talk" and negatively being followed by "sustain talk" may be understood by some examples from 7Cups. The following examples have been paraphrased using round-trip translation to ensure anonymity: \emph{'Life is hard' (Listener 1)} versus \emph{'Everyone is different. Some make friends quickly, some don't. It will take time to adjust. (Listener 2)}. From these examples of \emph{Reflection}, there is evidence that \emph{Reflection} varies in usage. \emph{Listener 1} will most likely contribute more to "sustain talk" as they are reflecting the member's current state of mind and not using \emph{Reflection} as a reframing tool. \emph{Listener 2}, conversely, reflects back the member's situation and reframes it in a more positive context, which is more likely to lead to "change talk". Therefore, we see from Table \ref{tab:RQ1_LR_Results} and these qualitative examples that our findings on \emph{Reflection} as a significant predictor of client satisfaction aligns well with prior research reporting \emph{Reflection} correlated with "change talk" and "sustain talk". 

The same prior work by Apodaca et al. \cite{MITI_change_talk} found that \emph{Affirm} was the only MI code predicted to be both significantly more likely to be followed by "change talk" and significantly less likely to be followed by "sustain talk" \cite{MITI_change_talk}. Our finding on \emph{Affirm} being a significant predictor of satisfaction builds upon this previous research to show that \emph{Affirm} is also well-received by clients within an online peer-counseling platform. \emph{Persuade (with permission)} is also found to be a significant predictor of success. While prior research on MI codes in offline mediums does not contradict this finding, this finding is also not as supported as the findings on \emph{Reflection} and \emph{Affirm}. Additionally, \emph{Persuade (with permission)} has a comparatively lower odds ratio of 1.019, indicating that for each added persuasive utterance, there is only $1.9\%$ increase in the odds of the conversation being satisfactory. This finding may be especially applicable to the development of MI usage in OMHCs. In the following examples: \emph{'I understand you must do what you think is best, but please try to eat something to stay healthy' (Listener 1)} versus \emph{'If I can make a suggestion, you should just ignore it' (Listener 2)}, \emph{Persuade (with permission)} clearly differs as \emph{Listener 1} persuades the member to make a change that will directly benefit their health, whereas \emph{Listener 2} persuades for a change that may or may not objectively benefit the member's health. These examples, combine with the statistically significant odds ratio, illustrate that \emph{Persuade (with permission)} may be an important component of satisfying conversations on OMHCs, if listeners are trained to use this MI code effectively.

In contrast to the MI-consistent positive predictors for client satisfaction, the MI-inconsistent code \emph{Inappropriate} has a significant negative correlation to satisfactory conversations. The significance of \emph{Inappropriate} as a negative predictor for client satisfaction highlights the need for listener training on OMHCs. Additionally, this finding aligns with previous research that MI-consistent techniques are more successful in promoting client language supporting behavioral change during counseling sessions than MI-inconsistent techniques \cite{MITI_change_talk, magill_MI_causal_model}. Of further note is that the MI-consistent codes \emph{Emphasizing Autonomy} and \emph{Closed Question}, while not statistically significant, have low $p$ values that are less than $0.1$. These findings offer additional evidence that the success of MI-consistent technique usage translates to client satisfaction within an online medium. 

\subsection{RQ2: How Does Listeners' Use of  Motivational Interviewing Techniques Change with Their Experience?}
\label{sec:RQ2}
Now we are interested in changes in listener usage of MI strategies over time. As such, we limit the analysis to those who have been active on the platform for at least a year. This gives us the ability to look at the changes in their use of MI techniques over time, unconfounded with differences between listeners who drop out quickly from those who persist on the site. We use the same analysis technique as Section \ref{sec:RQ1} to discover MI codes that show a change in usage as listeners gain more experience. 
We also examine the intersection of these identified codes with the codes from Section \ref{sec:RQ1} that were significantly predictive of client satisfaction.

First, we query listeners who have been active on the platform. We define active listeners in two ways: (1) the time between their last and first utterances should be at least one year and (2) they have participated in at least 500 sessions, because the average number of sessions per user satisfying the first criteria is 493 (which we round to the nearest hundred). The first method gives us a temporal filter while the second method allows us to discard those listeners who are inactive and only log onto the platform after a long time.
After identifying active listeners, we limit the analysis to conversations that contain at least 50 utterances to ensure that the discussion between the member and the listener moves beyond the introduction phase. After filtering the data, the average length of conversations is  713.6 utterances, with the largest containing  72,774 utterances. The relatively high average number of utterances per conversation when compared to our filter of 50 utterances shows that most conversations that exceed the threshold go on for many utterances. We use our automated labeling classifiers from Section \ref{sec:labeling} to find the MI techniques for the obtained filtered utterances. Correlational studies analysing the MI-Consistent behavior of active users is given in Appendix Subsection \ref{sec:rq2_appendix}. The studies describe the conversational level techniques used by listeners and generalization of behavior at the individualistic listener level.

\begin{figure}[t]

\includegraphics[ trim = {1.1cm 0cm 0.5cm 1.5cm},width=0.9\textwidth]{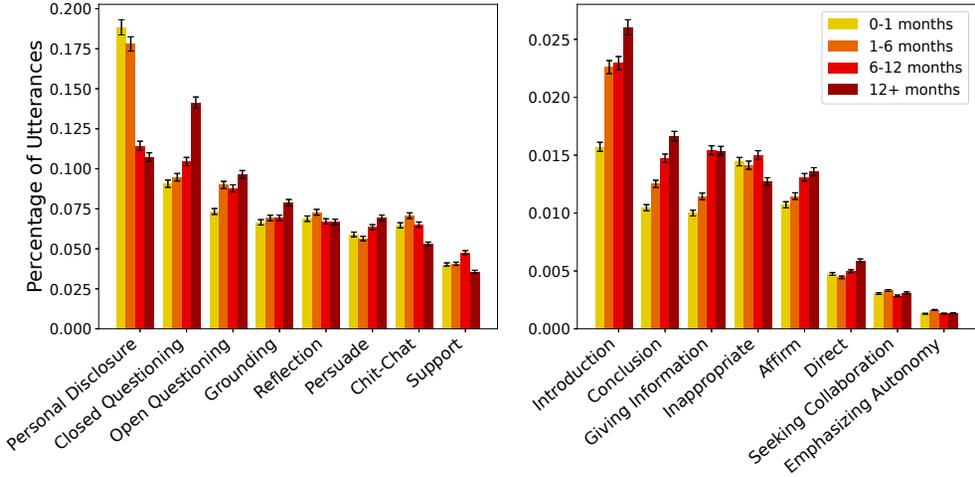}

    
    \vspace*{-4mm}
    \caption{Trends in Listener MI Codes over time.
    }
    \label{fig:rq2_trends}
\end{figure}

    

    

Figure \ref{fig:rq2_trends} presents the trends in usage of motivational interviewing codes as a listener's experience increases over time. We split the behavior of the listeners into four buckets: 0 to 1 months from joining, 1 to 6 months after joining, 6 to 12 months after joining, and more than 1 year after joining. This allows us to view trends in the platform as a user gains more experience. The $y$ axis of Figure \ref{fig:rq2_trends} shows the fraction of utterances for each bucket consisting of a MI technique. The left-hand graph in Figure \ref{fig:rq2_trends} shows changes in the relatively frequent MI codes (e.g., (\emph{Closed Question}, \emph{Open Questions}), \emph{Personal Disclosure}, \emph{Grounding}, \emph{Persuade}, \emph{Support}, \emph{Chit-Chat} and \emph{Reflection}), while the right-hand graph shows changes in the less frequent MI codes (e.g., \emph{Affirm}, \emph{Emphasizing Autonomy} and \emph{Seeking Collaboration}). The overall use of \emph{Affirm} is fairly low. Since Section \ref{sec:RQ1} showed that \emph{Reflection} and \emph{Affirmations} are the most useful MI technique for predicting conversation success based on Odds Ratio, these results suggest that training programs to encourage listeners to give more \emph{Reflections} could be helpful. We see increasing trends in \emph{Open Questions}, \emph{Closed Questions}, \emph{Introductions}, \emph{Conclusions}, \emph{Grounding}, \emph{Affirmations}, \emph{Persuade}, \emph{Giving Information} and \emph{Direct}. We also see decreasing trends in \emph{Personal Disclosure}, \emph{Reflection} and \emph{Inappropriate}. 

Increasing trends in \emph{Introductions}, \emph{Conclusions}, \emph{Closed Questions}, and \emph{Open Questions} show that listeners learn and refine a set method of conversing when they meet new support seekers. We also see more than 2.5$\%$ increase (given by the confidence intervals) in \emph{Direct} and \emph{Affirm}. This confirms that active and experienced users seem to learn to use more affirmations as they spend more time on peer-to-peer platforms. Decreasing trends are observed for \emph{Personal Disclosure} and \emph{Reflection}. This indicates that as listeners learn and understand the best ways to talk to support seekers, they rely less on sharing their own experiences (\emph{Personal Disclosure}) or summarizing the issues of the support seeker (\emph{Reflection}). The statistical significance of \emph{Reflection} for predicting client satisfaction seen in \ref{sec:RQ1} can be used to train listeners to learn this motivational interviewing technique. 

Change in listener MI usage can be  attributed to the listeners refining the way they communicate with members over time as they gain experiences. We observe this via a qualitative analysis on \emph{Reflection}, performed on randomly sampled active listeners by comparing reflective utterances within the first month of joining and after one year of experience as a listener. Within their first month of joining as a listener, we observe utterances like: \textit{'Haha doesn't everyone have that!' (Listener 1)}, \textit{'That isn't very helpful.' (Listener 2)}, and \textit{'You went to the adult side and it is a tough transition.' (Listener 3)}. These reflections are more repetitive of the member's utterance and add limited value to what members have said. On the other hand, after more than one year of experience as a listener, we see utterances which contain elements beyond paying attention; utterances which have exaggerated or amplified context, reflection of emotion/feeling that was not previously stated by member, and the use of analogies and comparatives.
For example: \textit{'I can see that he has a lot of narcissistic traits.' (Listener 1) [Amplified Context]}, \textit{'Well, we all need a safe place and that's it. It is okay to be stressed about love sometimes and feel like caged birds.' (Listener 2) [Analogy - Simile]}, and \textit{'I think it’s easy to feel like no one really cares.' (Listener 3) [Emotion]}. 

A qualitative examination of \emph{Personal Disclosure} shows that listeners learn to be more member-centric with experience. In the beginning months of joining, listeners disclosed their own emotions, intimate experiences and circumstances, to make support seekers more comfortable: \textit{'My parents are not supportive of college decisions.' (Listener 1)} and \textit{'Relationships are hard for me, my last partner did not wish to spend time with me.' (Listener 2)}. Alternatively, after spending more time on the platform, the listeners shared their mood, likes/ dislikes to make the member more comfortable: \textit{'I am feeling great today!' (Listener 1)} and \textit{'Fun fact, I am actually drinking tea right now. 10/10 I would suggest earl grey.' (Listener 2)}. Similarly, examination of \emph{Affirm} shows that listeners become more encouraging and positive as they gain more experience. For example, in the beginning months of joining, listeners only appreciate or acknowledge members' comments: \textit{'Thank you for sharing with me' (Listener 1)}, \textit{'That is an interesting perspective' (Listener 2)}. After gaining experience: \textit{'Those are amazing ideas, great work brainstorming designs' (Listener 1)}, \textit{'I know it is difficult to quit smoking, I am proud of you!' (Listener 2)}; we notice distinct usage of reinforcing and complementary words. The qualitative analysis of \emph{Affirm}, \emph{Reflections} and \emph{Personal Disclosures} shows that listeners adjust their interaction styles as they spend more time as a volunteer counselor.

\section{Conclusion}
This work provides a first-of-its-kind comprehensive data-set consisting of 14,797 chat messages annotated with motivational interviewing techniques on a peer-to-peer text-based volunteer counseling platform: 7Cups. Furthermore, this work translates the data into a supervised classification problem to automate the process of labeling, which helps us perform a large scale analysis of the data. Thereafter, the work focuses on a quantitative analysis which yields a few key takeaways:
\begin{itemize}
    \item \emph{Motivational interviewing techniques that lead to satisfactory conversations}: Our analysis shows that MI-consistent techniques are more successful in predicting client satisfaction. The techniques \emph{Reflection}, \emph{Affirm}, and \emph{Persuade} have the highest statistically significant positive correlations and odds ratios to satisfactory conversations. Additionally, we find that the MI-inconsistent technique \emph{Inappropriate} is a significant negative predictor for client satisfaction, pointing to the need for OMHCs to address inappropriate conversations in order to create better environments for both listeners and support seekers. 
    \item \emph{Change in behavior of listeners as they gain experience}: Active listeners learn to use the technique \emph{Affirm} more as they gain experience. Increasing trends in \emph{Introduction}, \emph{Conclusion}, \emph{Closed Question}, and \emph{Open Question} show that listeners learn and refine a set method of conversing when they meet new members. The study also shows that listeners talk less about themselves (\emph{Chit-Chat} and \emph{Personal Disclosure}), which reinforces previous findings that members gain more from conversations which are member-centric \cite{HAARD_CONVO_SUCCESS}. We note the decrease in the usage of \emph{Reflection} by listeners as they gain experience contrasts with its positive predictive ability of client satisfaction from Section \ref{sec:RQ1}. These findings on \emph{Reflection} can be used to build better training programs for listeners.
\end{itemize}

\subsection{Implications and Design Recommendation}
Our findings build on prior MI studies by exploring MI at scale within a unique peer-counseling online medium. Similar to previous works on MI and counseling success in offline settings \cite{MITI_change_talk, magill_MI_causal_model, effective_MI_1}, we also found that MI-consistent techniques were predictive of better outcomes (defined here as client satisfaction). Our work contributes back to these prior studies by showing that MI-consistent techniques translate well to online mediums with non-professional counselors. Due to the unique nature of the peer-counseling online medium of our study, our findings have several potential implications for practical applications within online platforms. 

Previous studies have shown that listeners on platforms like 7Cups feel that while the training program prepares them to help clients, it is not sufficient in training listeners to deliver high quality counseling sessions \cite{HAARD_CONVO_SUCCESS}.  Our work bridges this gap by providing a high-quality motivational interviewing dataset and fine-tuned language models which may be utilized for multiple cases. 

We recommend the use of the dataset for building suggestive models to empower the process of support exchange in online peer counseling platforms. Given a series of chats between the member and listener, artificial intelligence models could be made to do two things. First, the models can suggest to the listener a set of MI techniques to use. Second, chatbots can suggest utterances related to the suggested MI technique to listeners. This would help new listeners, who face the issue of inadequate training \cite{YAO_CMU} by guiding them to possible responses in various situations where member messages are extreme, irrational, and unreasonable. 
Listeners can be introduced to chatbots trained on member utterances to simulate real-life messaging with the diversity seen on peer-to-peer text-based counseling platforms. The released data can also be used for other annotations such as suicidal ideation detection. Other applications include detection of listeners who ask members for personal contact information, such listeners can be identified and flagged.

The models available can be used for real-time categorization of messages. Platforms could be integrated with inappropriate message classifiers to identify users sending obscene messages and hate speech. These users could be flagged for platform owners to review and potentially remove from the website. The classifier for \emph{Personal Disclosure} can be used to determine the first instance when the member feels comfortable enough to disclose their simple personal information about likes/dislikes, mood, and also for more intimate information about their issues. 

Lastly, the findings from our research questions give important insights into what MI techniques lead to satisfactory conversations within an online, text-based medium. Previous reports have indicated that current training programs are insufficient and leave counselors to develop their own strategies \cite{YAO_CMU}. As such, this novel research can be incorporated into OMHC training programs to improve counselor confidence and success. 
\subsection{Limitations}
This work has multiple limitations. The first limitation is that the work only covers one-to-one support seeker and peer counselor text-based platforms. It does not account for effects of group chat boards, or other Reddit-like online platforms. Second, this work does not capture the effects of offline activities, of the members and the listeners both. Many listeners connect with members on social media platforms despite being trained not to do so; this work is unable to capture interactions of users on alternate platforms. Platforms also allow volunteer counselors to seek support from other volunteer counselors, this work does not capture the impact of such interactions. It is possible that changes in MI techniques used by listeners is influenced due to the advice and guidance of other listeners. Inter-listener interactions and word-of-mouth advice may lead to drastic changes in listener behavior which are excluded from this study. 

Additionally, the majority of this study is quantitative in nature and needs to be augmented with qualitative interviews from users of such platforms. Furthermore, the F1 scores for the classifier models for \emph{Direct} and \emph{Seeking Permission} are around 0.5; these models may contain large amounts of false negatives due to their precision heavy nature. Another limitation is while the dataset provided may help platforms to build better training programs, it does not account for extreme situations of support seeker distress/behavior. Platforms like 7Cups often have support seekers who are unresponsive to suggestions and are unwilling to communicate in an appropriate manner. Well meaning peer counselors face enormous stress and anxiety while engaging such support seekers. This work excludes such chats from our analysis and released dataset to ensure that such support seekers do not deteriorate or relapse due to slight inaccuracy by our large language models. In such cases, peer counselors should redirect the support seekers to highly trained professional therapists and authority helpline numbers for further help.
\subsection{Generalization}
Our research examines motivational interviewing and client satisfaction on an OMHC. While our work is based in the specific domain of 7Cups, findings on the correlations between MI technique usage and client satisfaction may be more broadly applied to other online mental health groups.

\subsection{Ethical Considerations} This research has been approved by the Institutional Review Board (IRB) at our institution. All researchers involved in this study have signed strict confidentiality agreements with the 7Cups platform. As stated above, we received access to counseling session data from the platform 7Cups of Tea \cite{7cups} and worked with the 7Cups' Research Advisory Board. 7Cups follows HIPAA and confidentiality agreements when collecting user data. Additionally, due to the sensitive nature of this data, we took several precautions during annotation. Annotation was not outsourced, but rather was conducted solely by the researchers working on this project, each of whom conducted the Collaborative Institutional Training Initiative (CITI Program) training prior to annotation work. Messages were also anonymized by not including the usernames of members or listeners in sessions. All examples given in the paper are paraphrased from 7Cups using round-trip translation, instead of directly pulled, to further protect user confidentiality. Additionally, to protect the anonymity of user information within our dataset, we will release our dataset and language models (Mental Health Bert) subject to appropriate privacy and ethical considerations.

\section*{Acknowledgement}
We would like to thank the anonymous reviewers
for their helpful comments, and the team from 7 Cups for their feedback. This
work is funded in part by NSF AI Institute AI-CARING (IIS-2112633), and a NSF grant IIS-2144562. DY is supported by the Microsoft Faculty Fellowship.

\bibliographystyle{ACM-Reference-Format}
\bibliography{main}
\section{Appendix}
\subsection{Correlational analysis on the behaviors of active users}
\label{sec:rq2_appendix} 
\begin{figure}[!h]
    \includegraphics[trim={1cm 0cm 8cm 8.5cm}, clip, width=0.6\textwidth]{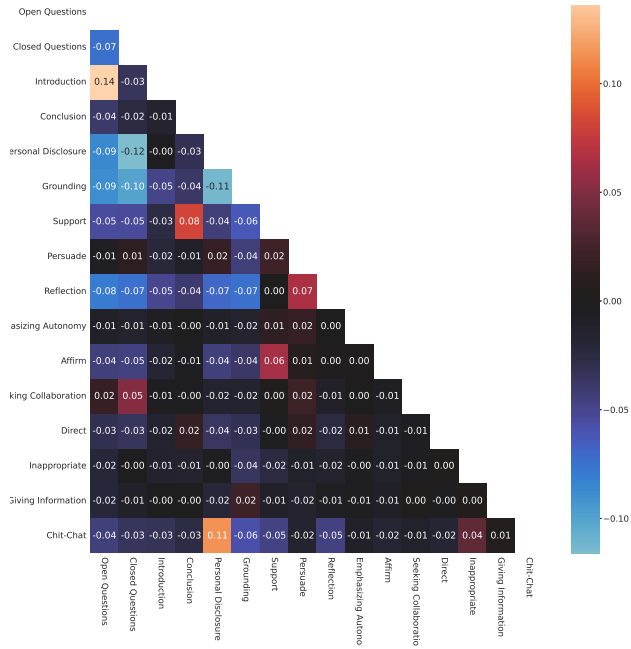}
    \caption{Correlation matrix for co-occurence of MI Codes}
    \label{fig:corr_mi_Code}
\end{figure}
We begin our correlational analysis on active users by finding which MI codes co-occur at an utterance level to understand listeners who are invested into online counseling platforms. The correlation matrix in figure \ref{fig:corr_mi_Code} is found on the multi hot encoding of the predicted codes. The overall correlation values in the figure are very low, although we observe noticeable positive correlations between \emph{Introduction} and \emph{Open Questions}, which is expected as discussed in Section \ref{sec:introduction}. We do not pursue this research direction to find the efficacy of multiple MI codes in the same utterance because users can send multiple messages in the same turn. Users treat an utterance as a unit of text which generally conveys one idea. The other reason lies in the fact that some MI codes are used much more than others, which pushes the correlation between codes towards negative values due to the multi hot encoding. This approach focuses on only the co-occurrence of MI codes, for our next analysis, we focus on the conversation level techniques used by our listeners. 
\begin{figure}[!h]
    \includegraphics[trim={1cm 0cm 8cm 8.5cm},clip, width= 0.6\textwidth]{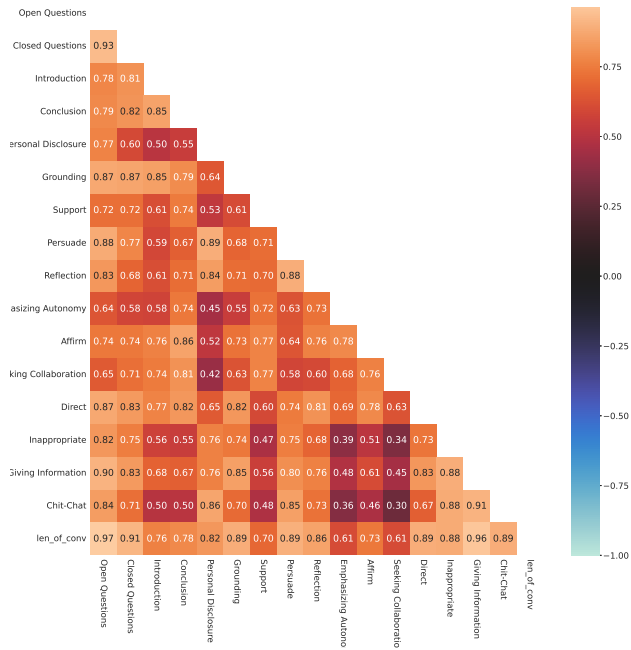}
    \caption{Co-relational Analysis at the Conversation Level}
    \label{fig:corr_conv_level}
\end{figure}
At a conversational level (figure \ref{fig:corr_conv_level}), we find that as the length of the conversation increases, the proportion of MI-consistent codes like \emph{Affirm}, \emph{Emphasizing Autonomy}, and \emph{Seeking Collaboration} decreases (i.e. as the length of the conversation increases, the counts of the MI-consistent codes does not increase with the same rate). The other finding is that \emph{Inappropriate} has some of the lowest correlation values in the entire matrix with all the other MI codes except for \emph{Giving Information}. This implies that active listeners are on the platform to spread positive messages.
The conversation level correlation analysis seems to show that \emph{Personal Disclosure} and \emph{Inappropriate} have lower co-relation with the MI-consistent codes like \emph{Affirm}, \emph{Seeking Collaboration}, \emph{Emphasizing Autonomy etc}. This indicates that listeners should not focus the conversations significantly on themselves.

\begin{figure}[!ht]
    \includegraphics[trim={0cm 0cm 8cm 8.5cm}, clip, width= 0.6\textwidth]{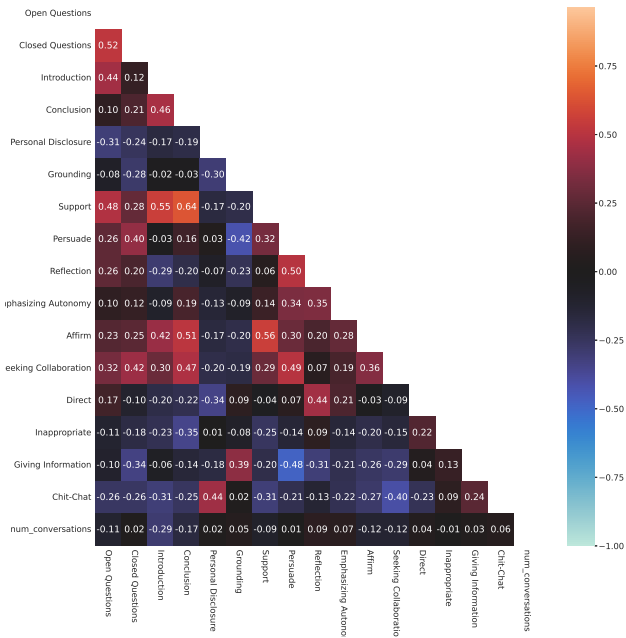}
    \caption{Co-relational Analysis at the Listener Level 
    }
    \label{fig:corr_list_level}
\end{figure}

The third type of correlational analysis focuses on the generalization of behaviors at the individualistic listener level. Co-relation between MI codes at the listener level would help us find possible relations such as a listener that sends more inappropriate messages may also send more messages to persuade the members. We find the co-relational matrix for the listener level in figure \ref{fig:corr_list_level}. We observe a high correlation between reflection and persuasion. This implies that listeners often repeat and reflect on what the member has uttered, and they also persuade the users to take up action items. Positive correlations between (\emph{Seeking Collaboration} and \emph{Conclusion}), and (\emph{Affirm} and \emph{Conclusion}) show that experienced listeners tend to take their conversations to its conclusion before signing off.
Other positive and negative listener level co-relation analysis show that in most cases experienced listeners follow MI-consistent behavior. \\~\\


\begin{table}[t]

\resizebox{0.9\columnwidth}{!}{%

\begin{tabular}{
>{\columncolor[HTML]{FFFFFF}}l 
>{\columncolor[HTML]{FFFFFF}}c 
>{\columncolor[HTML]{FFFFFF}}c 
>{\columncolor[HTML]{FFFFFF}}c 
>{\columncolor[HTML]{FFFFFF}}c 
>{\columncolor[HTML]{FFFFFF}}c 
>{\columncolor[HTML]{FFFFFF}}c }
\hline
\multicolumn{1}{c}{\cellcolor[HTML]{FFFFFF} Architectures: } &

  \multicolumn{3}{c}{\cellcolor[HTML]{FFFFFF}BERT-Base-Uncased} &
  \multicolumn{3}{c}{\cellcolor[HTML]{FFFFFF}BERT-large} \\ 
  \multicolumn{1}{c}{\cellcolor[HTML]{FFFFFF}} &
  \multicolumn{3}{c}{\cellcolor[HTML]{FFFFFF}(F1 Scores)} &
  \multicolumn{3}{c}{\cellcolor[HTML]{FFFFFF}(F1 Scores)} \\
  \hline
  \multicolumn{1}{c}{\cellcolor[HTML]{FFFFFF}MI Codes  $\backslash$ Context:} &
  \multicolumn{1}{p{2.5em}}{\cellcolor[HTML]{FFFFFF}None} &
  \multicolumn{1}{p{4.5em}}{\cellcolor[HTML]{FFFFFF}\small{1 previous utterance}} &
  \multicolumn{1}{p{4.5em}}{\cellcolor[HTML]{FFFFFF}\small{5 previous utterances}} &
  \multicolumn{1}{p{2.5em}}{\cellcolor[HTML]{FFFFFF}None} &
  \multicolumn{1}{p{4.5em}}{\cellcolor[HTML]{FFFFFF}\small{1 previous utterance}} &
  \multicolumn{1}{p{4.5em}}{\cellcolor[HTML]{FFFFFF}\small{5 previous utterances}}  \\ \hline
Giving Information &

  \multicolumn{1}{c}{\cellcolor[HTML]{FFFFFF}0.356} &
  \multicolumn{1}{c}{\cellcolor[HTML]{FFFFFF}0.329} &
  0.177 &
  
  \multicolumn{1}{c}{\cellcolor[HTML]{FFFFFF}0.422} &
  \multicolumn{1}{c}{\cellcolor[HTML]{FFFFFF}{\color[HTML]{212121} 0.371}} &
  0.318 
 \\
Reflection &
  \multicolumn{1}{c}{\cellcolor[HTML]{FFFFFF}0.567} &
  \multicolumn{1}{c}{\cellcolor[HTML]{FFFFFF}0.543} &
  0.608 &
  
  \multicolumn{1}{c}{\cellcolor[HTML]{FFFFFF}0.574} &
  \multicolumn{1}{c}{\cellcolor[HTML]{FFFFFF}0.607} &
  0.658 
 \\
Support &
  \multicolumn{1}{c}{\cellcolor[HTML]{FFFFFF}0.630} &
  \multicolumn{1}{c}{\cellcolor[HTML]{FFFFFF}0.633} &
  0.671 &
  
  \multicolumn{1}{c}{\cellcolor[HTML]{FFFFFF}0.608} &
  \multicolumn{1}{c}{\cellcolor[HTML]{FFFFFF}0.630} &
  0.711 
  
  \\
Affirm &

  \multicolumn{1}{c}{\cellcolor[HTML]{FFFFFF}0.485} &
  \multicolumn{1}{c}{\cellcolor[HTML]{FFFFFF}0.425} &
  0.333 &
  
  \multicolumn{1}{c}{\cellcolor[HTML]{FFFFFF}0.462} &
  \multicolumn{1}{c}{\cellcolor[HTML]{FFFFFF}0.397} &
  0.526 
 
  \\
Closed Question &

  \multicolumn{1}{c}{\cellcolor[HTML]{FFFFFF}0.894} &
  \multicolumn{1}{c}{\cellcolor[HTML]{FFFFFF}0.889} &
  0.858 &
  
  \multicolumn{1}{c}{\cellcolor[HTML]{FFFFFF}0.890} &
  \multicolumn{1}{c}{\cellcolor[HTML]{FFFFFF}0.877} &
  0.884

  \\
Persuade &

  \multicolumn{1}{c}{\cellcolor[HTML]{FFFFFF}0.595} &
  \multicolumn{1}{c}{\cellcolor[HTML]{FFFFFF}0.568} &
  0.649 &
  
  \multicolumn{1}{c}{\cellcolor[HTML]{FFFFFF}0.602} &
  \multicolumn{1}{c}{\cellcolor[HTML]{FFFFFF}0.740} &
  0.669

  \\
Open Questions &

  \multicolumn{1}{c}{\cellcolor[HTML]{FFFFFF}0.899} &
  \multicolumn{1}{c}{\cellcolor[HTML]{FFFFFF}0.904} &
  0.889 &
  
  \multicolumn{1}{c}{\cellcolor[HTML]{FFFFFF}0.904} &
  \multicolumn{1}{c}{\cellcolor[HTML]{FFFFFF}0.916} &
  0.904

  \\
Seeking Collaboration &

  \multicolumn{1}{c}{\cellcolor[HTML]{FFFFFF}\textbf{0.353}} &
  \multicolumn{1}{c}{\cellcolor[HTML]{FFFFFF}0.368} &
  NA &
  
  \multicolumn{1}{c}{\cellcolor[HTML]{FFFFFF}0.406} &
  \multicolumn{1}{c}{\cellcolor[HTML]{FFFFFF}0.388} &
  NA
  \\
Inappropriate* &

  \multicolumn{1}{c}{\cellcolor[HTML]{FFFFFF}0.342} &
  \multicolumn{1}{c}{\cellcolor[HTML]{FFFFFF}0.374} &
  0.261 &
  
  \multicolumn{1}{c}{\cellcolor[HTML]{FFFFFF}0.517} &
  \multicolumn{1}{c}{\cellcolor[HTML]{FFFFFF}0.454} &
  0.440\\
  
Direct &

  \multicolumn{1}{c}{\cellcolor[HTML]{FFFFFF}0.394} &
  \multicolumn{1}{c}{\cellcolor[HTML]{FFFFFF}0.311} &
  0.087 &
  
  \multicolumn{1}{c}{\cellcolor[HTML]{FFFFFF}0.410} &
  \multicolumn{1}{c}{\cellcolor[HTML]{FFFFFF}0.419} &
  0.182 \\
Emphasizing Autonomy &

  \multicolumn{1}{c}{\cellcolor[HTML]{FFFFFF}0.629} &
  \multicolumn{1}{c}{\cellcolor[HTML]{FFFFFF}0.621} &
  NA &
  
  \multicolumn{1}{c}{\cellcolor[HTML]{FFFFFF}0.694} &
  \multicolumn{1}{c}{\cellcolor[HTML]{FFFFFF}0.600} &
  NA 
  \\
Grounding &

  \multicolumn{1}{c}{\cellcolor[HTML]{FFFFFF}0.724} &
  \multicolumn{1}{c}{\cellcolor[HTML]{FFFFFF}0.743} &
  0.772 &
  \multicolumn{1}{c}{\cellcolor[HTML]{FFFFFF}0.749} &
  \multicolumn{1}{c}{\cellcolor[HTML]{FFFFFF}0.814} &
  0.803
  \\
Personal Disclosure &

  \multicolumn{1}{c}{\cellcolor[HTML]{FFFFFF}0.731} &
  \multicolumn{1}{c}{\cellcolor[HTML]{FFFFFF}0.719} &
  0.790 &

  \multicolumn{1}{c}{\cellcolor[HTML]{FFFFFF}0.771} &
  \multicolumn{1}{c}{\cellcolor[HTML]{FFFFFF}0.745} &
  0.804  \\
Introduction &

  \multicolumn{1}{c}{\cellcolor[HTML]{FFFFFF}0.854} &
  \multicolumn{1}{c}{\cellcolor[HTML]{FFFFFF}0.897} &
  0.730 &
  
  \multicolumn{1}{c}{\cellcolor[HTML]{FFFFFF}0.871} &
  \multicolumn{1}{c}{\cellcolor[HTML]{FFFFFF}0.894} &
  0.800

  \\
Conclusion &

  \multicolumn{1}{c}{\cellcolor[HTML]{FFFFFF}0.716} &
  \multicolumn{1}{c}{\cellcolor[HTML]{FFFFFF}0.729} &
  0.413 &

  \multicolumn{1}{c}{\cellcolor[HTML]{FFFFFF}0.701} &
  \multicolumn{1}{c}{\cellcolor[HTML]{FFFFFF}0.732} &
  0.465 \\
Chit-Chat &

  \multicolumn{1}{c}{\cellcolor[HTML]{FFFFFF}0.178} &
  \multicolumn{1}{c}{\cellcolor[HTML]{FFFFFF}0.619} &
  0.193 &
  
  \multicolumn{1}{c}{\cellcolor[HTML]{FFFFFF}0.293} &
  \multicolumn{1}{c}{\cellcolor[HTML]{FFFFFF}0.660} &
  0.240 
  
  \\
Other &

  \multicolumn{1}{c}{\cellcolor[HTML]{FFFFFF}0.345} &
  \multicolumn{1}{c}{\cellcolor[HTML]{FFFFFF}0.539} &
  0.667 &

  \multicolumn{1}{c}{\cellcolor[HTML]{FFFFFF}0.412} &
  \multicolumn{1}{c}{\cellcolor[HTML]{FFFFFF}0.622} &
  0.681
   \\
  \hline
  Avg F1 Scores (Across MI Codes)  & 0.570 & 0.600 & 0.539 & 0.605 & 0.639 & 0.608 \\
  \bottomrule
\end{tabular}%

}
  
  \caption{F1 scores of the Positive Classes for MI Codes using 0, 1, and 5 utterances as previous context when using BERT-Base-Uncased and BERT-Large architectures without domain specific pre-training.}
  \label{tab:f1scores_1}%
\end{table}

\begin{table}[b]

\resizebox{0.9\columnwidth}{!}{%

\begin{tabular}{
>{\columncolor[HTML]{FFFFFF}}l 
>{\columncolor[HTML]{FFFFFF}}c 
>{\columncolor[HTML]{FFFFFF}}c 
>{\columncolor[HTML]{FFFFFF}}c 
>{\columncolor[HTML]{FFFFFF}}c 
>{\columncolor[HTML]{FFFFFF}}c 
>{\columncolor[HTML]{FFFFFF}}c }
\hline
\multicolumn{1}{c}{\cellcolor[HTML]{FFFFFF} Architectures: } &

  \multicolumn{3}{c}{\cellcolor[HTML]{FFFFFF}BERTweet} &
  \multicolumn{3}{c}{\cellcolor[HTML]{FFFFFF}Roberta-Base} \\ 
  \multicolumn{1}{c}{\cellcolor[HTML]{FFFFFF}} &
  \multicolumn{3}{c}{\cellcolor[HTML]{FFFFFF}(F1 Scores)} &
  \multicolumn{3}{c}{\cellcolor[HTML]{FFFFFF}(F1 Scores)} \\
  \hline
  \multicolumn{1}{c}{\cellcolor[HTML]{FFFFFF}MI Codes  $\backslash$ Context:} &
  \multicolumn{1}{p{2.5em}}{\cellcolor[HTML]{FFFFFF}None} &
  \multicolumn{1}{p{4.5em}}{\cellcolor[HTML]{FFFFFF}\small{1 previous utterance}} &
  \multicolumn{1}{p{4.5em}}{\cellcolor[HTML]{FFFFFF}\small{5 previous utterances}} &
  \multicolumn{1}{p{2.5em}}{\cellcolor[HTML]{FFFFFF}None} &
  \multicolumn{1}{p{4.5em}}{\cellcolor[HTML]{FFFFFF}\small{1 previous utterance}} &
  \multicolumn{1}{p{4.5em}}{\cellcolor[HTML]{FFFFFF}\small{5 previous utterances}}  \\ \hline
Giving Information &

  \multicolumn{1}{c}{\cellcolor[HTML]{FFFFFF}0.422} &
  \multicolumn{1}{c}{\cellcolor[HTML]{FFFFFF}{ 0.517}} &
  0.233 
  &
  \multicolumn{1}{c}{\cellcolor[HTML]{FFFFFF}0.385} &
  \multicolumn{1}{c}{\cellcolor[HTML]{FFFFFF}0.410} &
  0.293 \\
Reflection &
  \multicolumn{1}{c}{\cellcolor[HTML]{FFFFFF}0.585} &
  \multicolumn{1}{c}{\cellcolor[HTML]{FFFFFF}0.705} &
  0.727 &
  
  \multicolumn{1}{c}{\cellcolor[HTML]{FFFFFF}0.580} &
  \multicolumn{1}{c}{\cellcolor[HTML]{FFFFFF}0.650} &
  0.649 \\
Support &
  
  \multicolumn{1}{c}{\cellcolor[HTML]{FFFFFF}0.692} &
  \multicolumn{1}{c}{\cellcolor[HTML]{FFFFFF}0.690} &
  0.775 
  &
  \multicolumn{1}{c}{\cellcolor[HTML]{FFFFFF}0.656} &
  \multicolumn{1}{c}{\cellcolor[HTML]{FFFFFF}0.651} &
  0.677 
  
  \\
Affirm &

  \multicolumn{1}{c}{\cellcolor[HTML]{FFFFFF}0.672} &
  \multicolumn{1}{c}{\cellcolor[HTML]{FFFFFF}0.542} &
  0.327 
  &
  \multicolumn{1}{c}{\cellcolor[HTML]{FFFFFF}{\color[HTML]{212121} 0.621}} &
  \multicolumn{1}{c}{\cellcolor[HTML]{FFFFFF}0.512} &
  0.387 
  \\
Closed Question &

  \multicolumn{1}{c}{\cellcolor[HTML]{FFFFFF}0.901} &
  \multicolumn{1}{c}{\cellcolor[HTML]{FFFFFF}0.901} &
  0.862 
  &
  \multicolumn{1}{c}{\cellcolor[HTML]{FFFFFF}0.896} &
  \multicolumn{1}{c}{\cellcolor[HTML]{FFFFFF}0.896} &
  0.894 
  
  \\
Persuade &

  \multicolumn{1}{c}{\cellcolor[HTML]{FFFFFF}0.605} &
  \multicolumn{1}{c}{\cellcolor[HTML]{FFFFFF}0.747} &
  0.816 &
  
  \multicolumn{1}{c}{\cellcolor[HTML]{FFFFFF}0.584} &
  \multicolumn{1}{c}{\cellcolor[HTML]{FFFFFF}0.616} &
  0.765 
  
  \\
Open Questions &

  \multicolumn{1}{c}{\cellcolor[HTML]{FFFFFF}0.911} &
  \multicolumn{1}{c}{\cellcolor[HTML]{FFFFFF}0.930} &
  0.921 
  &
  \multicolumn{1}{c}{\cellcolor[HTML]{FFFFFF}0.905} &
  \multicolumn{1}{c}{\cellcolor[HTML]{FFFFFF}0.915} &
  0.914 
  
  \\
Seeking Collaboration &

  \multicolumn{1}{c}{\cellcolor[HTML]{FFFFFF}0.512} &
  \multicolumn{1}{c}{\cellcolor[HTML]{FFFFFF}0.465} &
  NA
  &
  \multicolumn{1}{c}{\cellcolor[HTML]{FFFFFF}0.439} &
  \multicolumn{1}{c}{\cellcolor[HTML]{FFFFFF}{ 0.359}} &
  NA  \\
Inappropriate* &

  \multicolumn{1}{c}{\cellcolor[HTML]{FFFFFF}0.441} &
  \multicolumn{1}{c}{\cellcolor[HTML]{FFFFFF}0.595} &
  0.612 
  &
  \multicolumn{1}{c}{\cellcolor[HTML]{FFFFFF}0.412} &
  \multicolumn{1}{c}{\cellcolor[HTML]{FFFFFF}0.459} &
  0.462 \\
Direct &
  \multicolumn{1}{c}{\cellcolor[HTML]{FFFFFF}0.492} &
  \multicolumn{1}{c}{\cellcolor[HTML]{FFFFFF}0.500} &
  0.276 &
  \multicolumn{1}{c}{\cellcolor[HTML]{FFFFFF}0.415} &
  \multicolumn{1}{c}{\cellcolor[HTML]{FFFFFF}{\color[HTML]{212121} 0.400}} &
  0.182 \\
Emphasizing Autonomy &

  \multicolumn{1}{c}{\cellcolor[HTML]{FFFFFF}0.720} &
  \multicolumn{1}{c}{\cellcolor[HTML]{FFFFFF}0.667} &
  NA &
  \multicolumn{1}{c}{\cellcolor[HTML]{FFFFFF}0.769} &
  \multicolumn{1}{c}{\cellcolor[HTML]{FFFFFF}0.533} &
  NA \\
Grounding &

  \multicolumn{1}{c}{\cellcolor[HTML]{FFFFFF}0.776} &
  \multicolumn{1}{c}{\cellcolor[HTML]{FFFFFF}0.822} &
  0.828
  &
  \multicolumn{1}{c}{\cellcolor[HTML]{FFFFFF}0.738} &
  \multicolumn{1}{c}{\cellcolor[HTML]{FFFFFF}0.767} &
  0.810
  \\
Personal Disclosure &

  \multicolumn{1}{c}{\cellcolor[HTML]{FFFFFF}0.791} &
  \multicolumn{1}{c}{\cellcolor[HTML]{FFFFFF}0.755} &
  0.827 &
  
  \multicolumn{1}{c}{\cellcolor[HTML]{FFFFFF}0.776} &
  \multicolumn{1}{c}{\cellcolor[HTML]{FFFFFF}0.741} &
  0.800 \\
Introduction &

    \multicolumn{1}{c}{\cellcolor[HTML]{FFFFFF}0.895} &
  \multicolumn{1}{c}{\cellcolor[HTML]{FFFFFF}0.908} &
  0.841
  &
  
  \multicolumn{1}{c}{\cellcolor[HTML]{FFFFFF}0.855} &
  \multicolumn{1}{c}{\cellcolor[HTML]{FFFFFF}0.893} &
  0.828
  
  \\
Conclusion &

  \multicolumn{1}{c}{\cellcolor[HTML]{FFFFFF}0.798} &
  \multicolumn{1}{c}{\cellcolor[HTML]{FFFFFF}0.748} &
  0.521 &
  
  \multicolumn{1}{c}{\cellcolor[HTML]{FFFFFF}0.739} &
  \multicolumn{1}{c}{\cellcolor[HTML]{FFFFFF}0.710} &
  0.489 \\
Chit-Chat &

  \multicolumn{1}{c}{\cellcolor[HTML]{FFFFFF}0.269} &
  \multicolumn{1}{c}{\cellcolor[HTML]{FFFFFF}0.680} &
  0.397 
  &
  \multicolumn{1}{c}{\cellcolor[HTML]{FFFFFF}0.253} &
  \multicolumn{1}{c}{\cellcolor[HTML]{FFFFFF}0.604} &
  0.214 
  \\
Other &

  \multicolumn{1}{c}{\cellcolor[HTML]{FFFFFF}0.446} &
  \multicolumn{1}{c}{\cellcolor[HTML]{FFFFFF}0.623} &
  0.769
  &
  
  \multicolumn{1}{c}{\cellcolor[HTML]{FFFFFF}0.310} &
  \multicolumn{1}{c}{\cellcolor[HTML]{FFFFFF}0.567} &
  0.755 \\
  \hline
  Avg F1 Scores & 0.642 & 0.693 & 0.649 & 0.608 & 0.628 & 0.606\\
  (Across MI Codes) &  & &  &  & & \\
  \bottomrule
\end{tabular}%
}
  
  \caption{F1 scores of the Positive Classes for MI Codes using 0, 1, and 5 utterances as previous context when using vinai/BERTweet and Roberta-Base architectures without domain specific pre-training.}
  \label{tab:f1scores}%
\end{table}

\begin{table*}[!h]

  \begin{tabular}{p{6cm}p{2.5cm}p{4cm}}
     \toprule
    Class Name (Abbreviation) & Inter Annotator Agreement &  Agreement Between the model and the Annotators \\
    \midrule
    Giving Information (GI) & 1 & 0.663\\
    Reflection (RF) & 0.587 & 0.552\\
    Support	(SUP) &0.710 & 0.594\\
    Affirm (AFFIRM)& 1 & 0.795\\
    Closed Question	(QUC) & 0.826 & 0.826\\
    Open Question (QUO) &0.808 & 0.729 \\
    Persuade (PR) & 0.581 & 0.47 \\
    Direct (DI) & 1 & 0.663\\
    Emphasizing Autonomy (AUTO)& 1 & 1\\
    Grounding (GR) &0.906 & 0.796\\
    Personal Disclosure (PD)& 0.594 & 0.499\\
    Introduction/Greeting (INT)& 1 & 0.795\\
    Outro/Conclusion (OUT)& 0.795 & 0.795\\
    Seeking Collaboration/Permission (SEEK) & 1 & 0.663\\
    Inappropriate (IP)& NAN & NAN\\
    Chit-Chat (CC)& 0.71 & 0.426 \\
    \hline
    Cumulative Agreement & 0.734 & 0.614 \\
    \bottomrule
  \end{tabular}
  \caption{Validating the Binary classifiers by calculating Krippendorff Alphas for between the Annotators and the MH-BERT Model.}
  \label{tab:annotation_agreement_validation}
\end{table*}

\end{document}